# Synthesis methods of graphitic carbon nitride: a superior photocatalyst

**SOUMIK KUMAR KUNDU[1], SAMIT KARMAKAR[1] and G. S. TAKI[1,*]**

[1] Institute of Engineering & Management, Salt Lake, Kolkata-700091, India
*Author for correspondence (gstaki@iemcal.com)

**Abstract.** In recent years, conjugated polymers such as graphitic Carbon Nitride (g-$C_3N_4$) attracts major attention to the researchers for the harnessing of renewable energy and environmental remediation through photocatalytic water splitting. Its moderate electronic band gap structure helps to absorb large spectrum of abundant solar radiation for the generation of hydrogen, a high density chemical energy source, by water splitting method. Its outstanding physicochemical stability makes it a reliable energy conversion material. Another key attribute to the researchers is the simple way of synthesizing pristine g-$C_3N_4$ and its nanocomposite structures modified with metallic and non-metallic materials. g-$C_3N_4$ can be synthesized in both chemical and physical process. In this work, the superiority in structural, optical and photocatalytic property observed in physically developed g-$C_3N_4$ over chemically synthesized g-$C_3N_4$ has been discussed and based on such studies, a suitable synthesis method has been proposed.

**Keywords.** Graphitic carbon nitride; conjugated polymer; renewable energy; photocatalytic water splitting; physicochemical stability.

## 1. Introduction

A large amount of energy per capita is spent to maintain modern lifestyle. Fossil fuels such as Petroleum, Coal, Natural gases are the main sources of energy which has limited resources. Moreover, burning of such fuels causes harmful emission that degrades the ecosystem. The researches on renewable ecofriendly energy harnessing and environmental remediation is essential now a days. Globally an intensive research has been continued to convert the abundant solar energy to chemical and electrical energy. Hydrogen is an important chemical energy source due to its high energy density and its consumption has no harmful emission.

Among several methods of Hydrogen generation, the photocatalytic water splitting is an ecofriendly efficient method. At a large extent, this process is identical to natural photosynthesis occurring in plants' kingdom. In such artificial photosynthesis [1-4], chlorophyll is replaced by the photocatalysts. Generally, materials with semiconducting properties such as Titania ($TiO_2$), Zinc Oxide (ZnO), Hematite (α-$Fe_2O_3$), Cadmium Sulfide (CdS), Strontium titanate ($SrTiO_3$), Cadmium Selenide (CdSe), Lead Sulfide (PbS) etc. are used as photocatalysts. In photocatalytic water splitting process, due to the incidence of photons on photocatalytic materials, valence band electrons ($e^-$) get energized and reaches to the conduction band edge. Simultaneously, holes ($h^+$) are left out at the valence band edge. As a result, this charge separation creates a potential difference which is essential for water splitting. Oxygen releases due to water splitting in presence of holes in the valence band. Hydrogen ions get reduced by photo generated electrons, in the conduction band region, to produce hydrogen gas. The oxidation-reduction reaction is shown here.

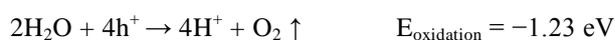

$2H_2O + 4h^+ \rightarrow 4H^+ + O_2 \uparrow \qquad E_{oxidation} = -1.23$ eV

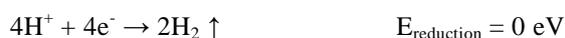

$4H^+ + 4e^- \rightarrow 2H_2 \uparrow \qquad E_{reduction} = 0$ eV

It has been observed from the equation that minimum potential required for photocatalytic water splitting is 1.23 V. A photocatalyst possessing 1.23 eV bandgap energy can develop that amount of potential for water splitting. But low bandgap enhances the chances of charge carrier recombination which decreases both the charge separation and ability of developing water splitting potential. The sunlight at earth surface possesses around 4% ultra violet (<400nm), ~42% visible (400–700nm) and nearly 54% infrared (>700nm) light energy. The photocatalyst with large bandgap (>3.5 eV) can separate the charge carriers very efficiently by absorbing only UV light energy which eventually reduces the photo conversion efficiency. So, a photocatalyst with moderate bandgap energy is preferred that can separate these charge carriers to develop water splitting potential utilizing broad spectrum of solar radiation. This work highlights a comparative study of various photocatalysts along with a detail study of graphitic carbon nitride produced in popular methods.

## 2. Various photocatalysts

Usually, the semiconducting materials are used for photocatalytic applications. The photocatalytic water splitting process using semiconductor was pioneered by Fujishima and Honda in 1972 [5]. They used $TiO_2$ as a photocatalyst. Since then, researchers around the globe have utilized this material in both pristine and modified form and achieved moderately good results. This material also showed good physicochemical stability for a lengthy period of time, even during degradation of organic pollutants. This property of $TiO_2$, supports the stable environmental remediation [6-15]. But its high electronic bandgap ~3.3 eV allows it to absorb mainly UV and a part of visible spectrum of solar energy that lowers its overall energy conversion efficiency. Composite structures of $TiO_2$ have solved this problem to a certain extent but it introduces toxicity in the system. Some researchers have studied ZnO nanostructures for its interesting atomic structure that can absorb a small part of visible range of solar radiation. But its photocorrosion due to UV irradiation, deteriorates its stability for long time effective use. Pristine and Composite hetero-structures of ZnO have achieved some acceptable results [16-27]. Besides these two well-known materials, few researchers have studied various nanostructures of α-$Fe_2O_3$, $SrTiO_3$, CdS, CdSe, PbS etc. and achieved slightly improved results [28-33]. Various photocatalysts with their different bandgap energies and their probable photo absorption range from the spectrum of solar light is shown in the figure. (see figure 1).

In the recent years, the semiconducting materials e.g. $TiO_2$, ZnO, $SrTiO_3$, α-$Fe_2O_3$ etc. are widely used in photocatalytic applications. But their electronic band structures are not very much suitable for efficient energy conversion. Moreover, these photocatalysts show poor stability and high toxicity. The conjugated polymer g-$C_3N_4$ has drawn the attraction to the current researchers as a special photocatalyst. g-$C_3N_4$ has stable existence possessing a moderate bandgap energy among the other allotropes of carbon nitrides. The phase g-$C_3N_4$ is composed of the basic structural unit of triazine or heptazine (tri-s-triazine) ring is shown in the figure (see figure 2). The selection of basic unit solely depends upon the synthesis process temperature. The g-$C_3N_4$ act as an n-type semiconductor due to presence of free electrons at valence band edge. Free electrons are generated due to the occurrence of $sp^2$ C–N bonding. The $sp^2$ hybridized C atom has 3 $sp^2$ hybrid orbitals and a p-orbital with single electron at each orbital. $sp^2$ hybridized N atom has similar number of $sp^2$ orbitals and p-orbitals but it has 1 $sp^2$ orbital with paired electrons. In a triazine unit of g-$C_3N_4$, double bonds and single bonds of carbon and nitrogen atoms are present in alternate fashion. During double bond formation between C and N, 1 $sp^2$ hybrid orbital of each atom overlaps head to head to form σ-bond and p-orbitals overlap in parallel to form л-bond. In single bond formation only a σ-bond is formed between C and N atoms. In the triazine structure the unused $sp^2$ orbital electrons (lone pairs) of N atoms are the free electrons that enhances its electrical conductivity. These free electrons also enhance its optical properties by forming a bandgap energy of ~2.7 eV between VB and CB (see figure 3). At such bandgap it can absorb visible range solar irradiation which eventually increases photo-conversion efficiency. The N atom also helps to stabilize the structure at ambient environment [34,35].

## 3. Property of graphitic Carbon Nitride

Studying the structural and optical property of a catalyst is extremely important for hydrogen generation by photocatalytic water splitting. The photoluminescence (PL) is an optical property of such photocatalyst that occurs due to the charge carrier recombination. After absorption of photon energy, the electrons in the valence band get excited and reaches to the conduction band edge. At the time of attaining the stable state at valence band edge they release some photon energy in form of light. This phenomenon is called photoluminescence. An idea of bandgap energy of a photocatalyst can be obtained from the photo emission peak of PL spectra. Reduction of photo emission peak signifies good charge carrier separation which is essential for an efficient photocatalyst. The enhancement of these properties are very much synthesis process dependent which is also discussed here.

### 3.1 *Property of g-$C_3N_4$ synthesized in chemical route*

The g-$C_3N_4$ synthesis was pioneered by Xinchen Wang *et al.* in 2008 [36]. They have synthesized g-$C_3N_4$ by thermal poly-condensation method of cyanamide and observed that absorption of solar energy spectra shifts towards visible range (red shift) with increasing temperature of poly-condensation from ~400°C to 600°C. In 2014, Fang He *et al.* synthesized a sheet on sheet composite structure of g-Poly Acrylo Nitrile (PAN)/g-$C_3N_4$ by direct heating of melamine mixed with PAN [37]. They observed enhanced photo absorption and decreased PL intensity with increasing wt% of g-PAN in the g-PAN/g-$C_3N_4$ composite. 10 wt% g-PAN/g-$C_3N_4$ shows the lowest PL intensity. The chemically synthesized pristine g-$C_3N_4$ needs metal doping to enhance photo absorption spectra towards visible range. Few researchers used noble metal sensitization for this purpose. Yanfeng Chen *et al.* synthesized a heterostructure of g-$C_3N_4$/Ag/$TiO_2$ microsphere. It has been observed that PL intensity of protonated g-$C_3N_4$ was very high and it decreased significantly after creating heterostructure with $TiO_2$. The lowest PL intensity was observed in g-$C_3N_4$(4%)/Ag/$TiO_2$ heterostructure (see figure 4) [38]. In 2015, Jingrun Ran *et al.* developed porous phosphorus doped g-$C_3N_4$ nanosheet by thermal exfoliation method of combined porous phosphorus and g-$C_3N_4$. They achieved better absorption range after doping with non-metal porous phosphorus. It resulted in good photocatalytic $H_2$ production [39]. The metal doping enhances absorption range with an increase in PL intensity due to the recombination of photo generated electrons and holes. High PL intensity degrades the quality of a photocatalyst. To decrease the PL intensity, some researchers doped metal and non-metal with the nano-composite structure of g-$C_3N_4$. Considering this attribute, Xiaoling Ding *et al.* synthesized heterostructure of CdS/Au/g-$C_3N_4$. The gold nano particles were reported to enhances both visible light absorption and PL intensity. CdS doping resulted in reduced PL intensity and enhanced overall photocatalytic $H_2$ production [40]. Shijing Liang *et al.* observed that light absorption edge shifted towards visible range after chemically doped Au on g-$C_3N_4$. They also observed high PL intensity at a broad spectrum ranging from ~420 nm to 500 nm. The intensity was reduced after Pt sensitization on the Au/g-$C_3N_4$ composite (see figure 5) [41]. Sayan Bayan *et al.* demonstrated the enhancement of surface plasmon property along with photoluminescence tuning of g-$C_3N_4$ by attaching gold nanoparticle to it [42]. K. K. Chattopadhyay *et al.* studied PL intensity that can be tuned by inducing contamination in precursor material during the time of synthesis [43]. Yidan Luo *et al.* synthesized non-metal elements (Boron, Phosphorus, Fluorine) doped g-$C_3N_4$ in chemical method to study its optical property. They have observed Fluorine doped g-$C_3N_4$ possesses highest absorption range and lowest PL intensity peak [44]. PL intensity can also be decreased by reducing the graphitic carbon nitride. Samit Kumar Ray *et al.* achieved low PL intensity after reduction of the synthesized g-$C_3N_4$ [45].

3.2 *Property of g-$C_3N_4$ synthesized in physical route*

Leilei Guan *et al.* synthesized g-$C_3N_4$ nanocone (g-CNNC) array by DC discharge plasma sputter reaction deposition method. They utilized $CH_4$, $N_2$ and $H_2$ gas mixture as a precursor for the development of g-$C_3N_4$ nanostructure. They varied the precursor gas mixing ratio $CH_4$:($N_2$+$H_2$) ranging from 1/10–1/150 to study its structural transition from silicon nanocone to g-CNNC and optical property [46-48]. They have observed that at 1/10 and 1/20 ratio a perfect conical shape of silicon nanocone with 2-3μm height and 20°–25° cone angle was formed. As the ratio was decreased to 1/40, a coarse structure of g-$C_3N_4$ is formed. When the ratio was 1/60, the structure was improving. The size of g-CNNC became progressively larger with further decreasing the ratio from 1/60–1/100. At 1/150 ratio the biggest and highest g-CN nanocone was formed. They have studied photoluminescence property by exciting the samples at 325 nm light source. It was observed that, no peak was found when methane wasn't injected which ensures it was a diamond nanocone. Because bandgap of diamond crystal is 5.5 eV which cannot be excited by 325 nm light. There was no peak when the $CH_4$:($N_2$+$H_2$) ratio was 1/20 and above, which signified the formation of silicon nanocone. At 1/40 $CH_4$:($N_2$+$H_2$) ratio, a wide peak centered at 440 nm was observed which is attributed to the tri-s-triazine based g-$C_3N_4$ planar structure. When $CH_4$:($N_2$+$H_2$) ratio decreased to 1/100, the peak height increased and FWHM became narrow, implied improved crystallization but poor optical property. At $CH_4$:($N_2$+$H_2$) ratio of 1/150 a strong PL intensity was observed at 454 nm also illustrated as superior crystallinity but worst optical property (see figure 6).

4. **Synthesis of graphitic Carbon Nitride**

4.1 *Chemical Process:* The g-$C_3N_4$ nanostructure synthesis was pioneered by Xinchen Wang *et al.* [36]. They have used chemical route for synthesis. In chemical approach, the graphitic-$C_3N_4$ nanostructure is synthesized by thermal polymerization of nitrogen enriched precursors such as cyanamide, melamine, urea, thiourea etc. (see figure 7) Thermal polymerization is the combination of poly-addition and poly-condensation. With increasing temperature, the precursors are usually condensed as their higher forms and at ~550°C temperature finally the tri-s-triazine based g-$C_3N_4$ is formed.

4.2 *Physical Process:* The synthesis of graphitic-$C_3N_4$ nanostructure by sputtering method was performed by Leilei Guan et. al. They synthesized g-$C_3N_4$ nanocone (g-CNNC) array by DC plasma sputtering process [46]. The Si substrate was polished with diamond powder. It was a two-step process. In first step, 100 nm thick Ni catalyst was deposited on polished Si substrate by ablating Ni target using Nd:YAG laser source under 7.5 μtorr vacuum pressure. In the second step, the Ni coated Si substrate was put on a graphite supporter as cathode, 0.5 cm below the conical shaped anode. The precursor gas mixture of $CH_4$:($N_2$+$H_2$) was inserted into the chamber maintaining a total pressure of 5.62 torr. A DC plasma was generated above the substrate through DC glow discharge. In the DC discharge plasma, precursor mixture was dissociated and $\dot{C}H_x$ radicals, $H^+$ ions and $N^+$ ions were formed. Here, $\dot{C}H_x$ radicals and $H^+$ ions had an important role for generation of g-CNNC. $CH_4$ did not contribute its C atom to form CN structure. The main role of $H^+$ ion is to sputter graphite frame to generate C atoms for development of g-CNNC structure and another role is to etch graphite and amorphous carbon structure grown on Si substrate. In absence of $\dot{C}H_x$ radicals a large amount of C atoms were generated by $H^+$ ion sputtering which eventually develop Diamond nanocone (DNC). The bond dissociation energy of $H_2$ atom and $\dot{C}H_x$ radicals are nearly same (4.3-4.6 eV). So, in presence of $\dot{C}H_x$ radicals, they react with H atoms to form $CH_4$, maintaining a ratio of ~0.75 between C atom and N atom to generate $C_3N_4$ structure.

4.3 *Proposed Process:* A synthesis process has been proposed here with a view to develop an optically enhanced pure g-$C_3N_4$ nano structure for photocatalytic application. It has been planned to use a double-head DC magnetron sputtering setup for g-$C_3N_4$ synthesis. The g-$C_3N_4$ nanostructure will be developed on a transition metal coated Si substrate by DC magnetron sputtering setup under a fine vacuum condition. Here, transition metal e.g., Ni, Cu, Ti etc. will be utilized as target material in one of the two magnetron sputter-head. Argon gas will be injected into the sputtering chamber to produce DC argon plasma. In presence of magnetic field, the confined electrons sustain the stable DC discharge. The target atoms are sputtered out due to the bombardment of $Ar^+$ ion and proceed towards the Si substrate kept at a suitable distance from the target. The film thickness of the transition metal would be optimized by varying the discharge voltage or current, substrate biasing voltage and distance between substrate and target. After transition metal deposition, the second magnetron sputtering head containing graphite target plate will be used for g-$C_3N_4$ synthesis. Only a gas mixture of $N_2+H_2$ maintaining a certain ratio will be injected through the magnetron head to produce a mixed $N_2+H_2$ plasma. Molecular carbon from the graphite target would be sputtered out due to the bombardment of energetic $H^+$ ion. Here, sputtered carbon atoms will be ionized and ions will propagate towards the substrate. During passing through the plasma, C atoms will interact with the N atoms to form carbon nitride compound. The energy required for producing g-$C_3N_4$ will be optimized by varying both the target potential and substrate bias potential. The resultant enhanced production of g-$C_3N_4$ will be adjusted by varying the gas mixing ratio, operating vacuum and the distance from the target to substrate. Here $CH_4$ gas may not be needed to use due to the fact that $CH_4$ react with $N_2$ atoms will produce CN radicals that opposes the formation of crystalline g-$C_3N_4$ [48]. Moreover, the presence of excess $H^+$ ion formed due to the dissociation of $CH_4$, causes sputtering of the grown surface. The schematic view of the whole process is shown in figure (see figure 8).

5.  Conclusion

The present study on the synthesis of g-$C_3N_4$ in chemical and physical route shows that g-$C_3N_4$ synthesized in thermal polymerization process of chemical route possess a few undesired functional groups which are very difficult to remove. Such functional groups are absent in physical synthesis process. Moreover, further processing of the basic granular product is essential to prepare a useful structure. A useful structure is possible to prepare in physical route in a single step under clean vacuum condition. The functional groups remain after the chemical process may also damage the $sp^2$ CN bond and diminishes the optical property by eliminating free electrons. To compensate such problems, the metallic and non-metallic materials are needed to dope with chemically synthesized g-$C_3N_4$ to increase free electrons that enhance its optical property by reducing bandgap. The desired bandgap can be tuned by metallic and non-metallic doping. Hence, a lengthy process consists of several steps is essential to prepare a useful composite structure. On the other hand, in proposed magnetron sputtering process operating under a fine vacuum condition maintains the purity of the sample and a suitable integrated stable structure may be prepared by optimizing gas mixing ratio of $N_2:H_2$, discharge intensity and substrate biasing voltage. The distance between target and substrate along with the process time will also be optimized. The dense plasma generated near the vicinity of magnetron sputtering head can provide the essential thermal condition for the synthesis. Large external heating needed for the poly-condensation method is absent in the proposed method. A DC magnetron sputtering setup, consisting of R.F. and DC substrate biasing facility, has already been indigenously developed with a view to synthesize carbon nitride nanostructures at Applied Materials Research Laboratory (AMRL), Dept. of E.C.E., IEM, Kolkata. Facility is also provided to tune discharge intensity, sputtering yield and gas mixing ratio in the reactor.

# Figures

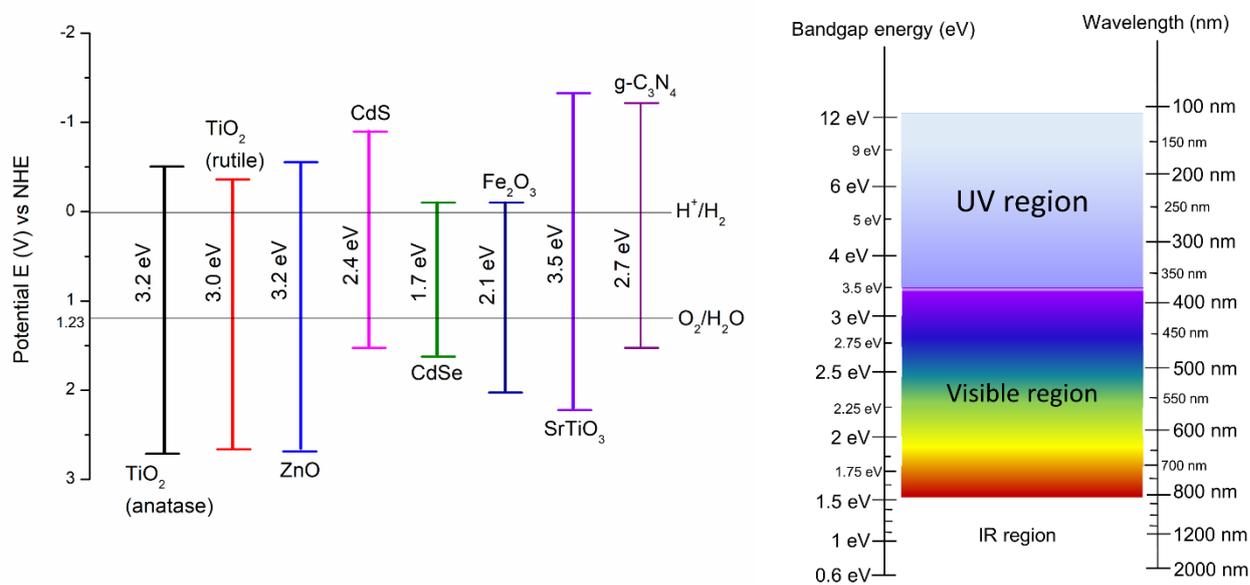

**Figure 1.** Various photocatalysts with their different bandgap energies and spectrum of solar light energy

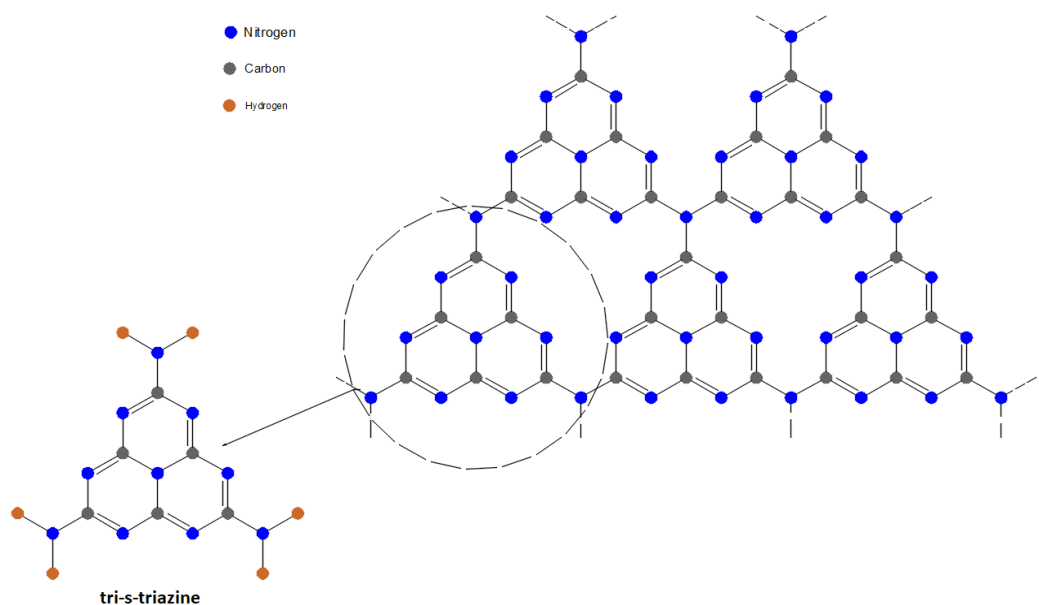

**Figure 2.** tri-s-triazine based g-$C_3N_4$ structure

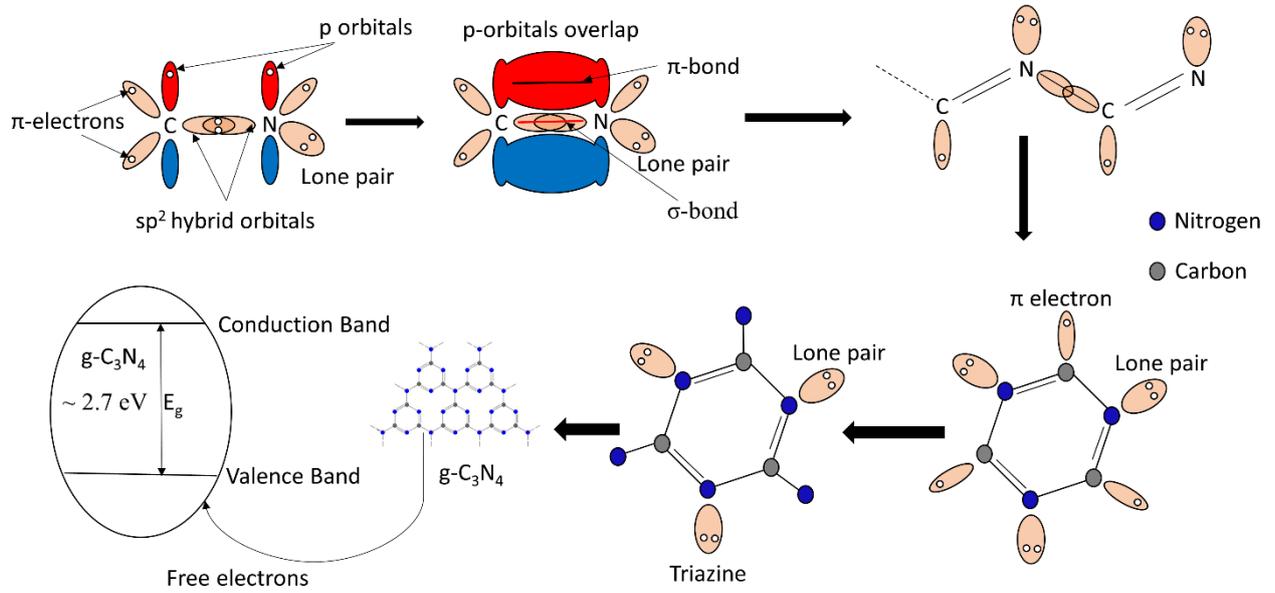

**Figure 3.** Schematic view of sp$^2$ C–N bond formation and free electron generation

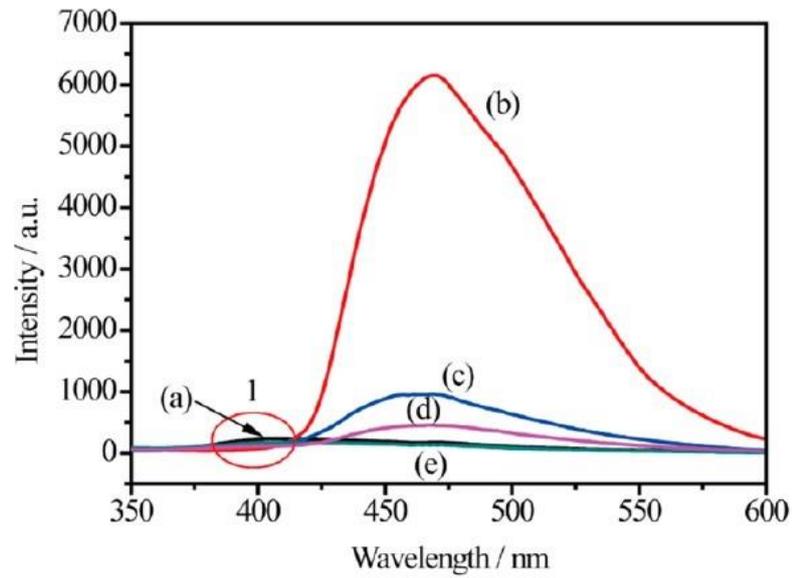

**Figure 4.** PL spectra of (a) TiO$_2$ microspheres, (b) protonated g-C$_3$N$_4$, (c) TiO$_2$/g-C$_3$N$_4$ microspheres, (d) Ag/TiO$_2$ microspheres, (e) g-C$_3$N$_4$ (4%)/Ag/TiO$_2$ ; as shown by Y. Chen et al. [38]

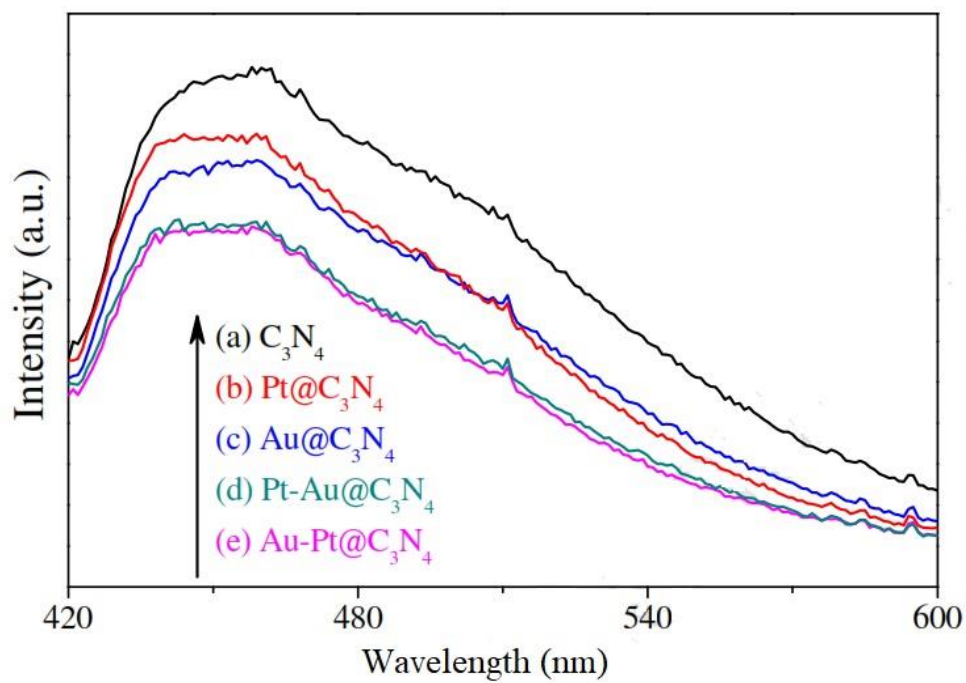

**Figure 5.**    Photoluminescence spectra under 400 nm excitation at 298 K; as shown by S. Liang et al. [41]

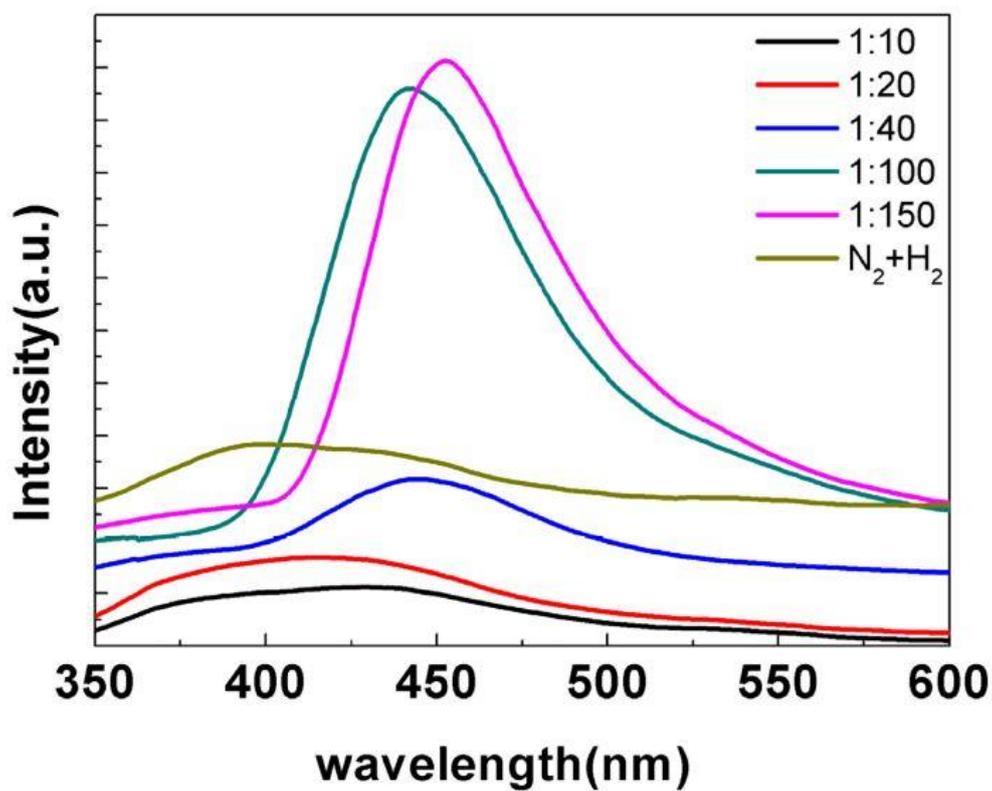

**Figure 6.**    PL spectra of the nanocone arrays grown with different $CH_4/(N_2+H_2)$ ratios of 1/10, 1/20, 1/40, 1/100, 1/150 and 0 respectively as shown by L. Guan et al. [46].

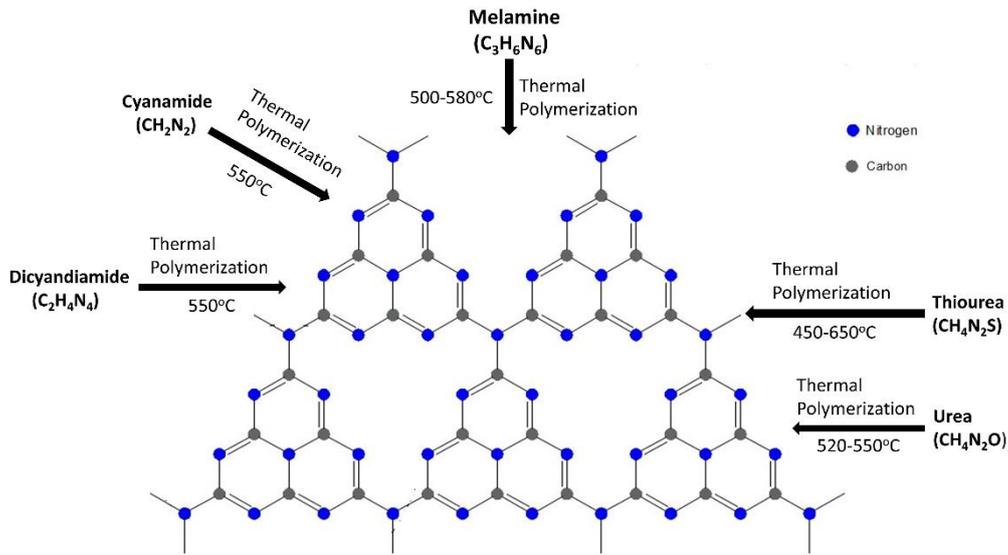

**Figure 7.** Schematic illustration of g-$C_3N_4$ synthesis by thermal polymerization method.

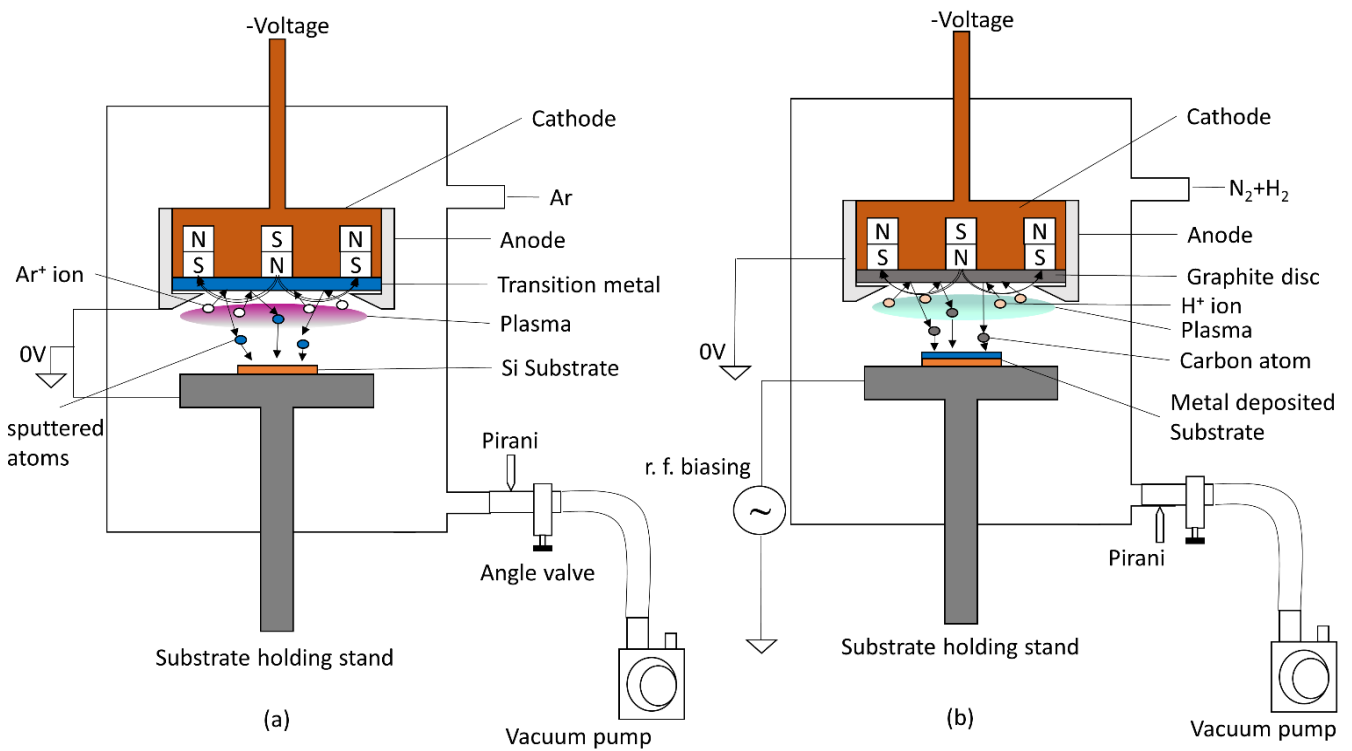

**Figure 8.** Schematic view of the proposed g-$C_3N_4$ synthesis process (a) Transition metal sputtering deposition on Si substrate (b) g-$C_3N_4$ synthesis on transition metal coated substrate